\documentclass[graybox]{svmult}

\usepackage{graphics} 
\usepackage{epsfig} 
\usepackage{amsmath} 
\usepackage{amssymb}  
\usepackage{subfig}
\usepackage{array}
\usepackage{booktabs}
\usepackage[export]{adjustbox}
\usepackage{paralist}
\usepackage{commath}

\usepackage{color}
\usepackage{booktabs}
\usepackage{flushend}
\usepackage{multirow}
\usepackage{enumitem}
\usepackage{longtable}
\usepackage{afterpage}

\usepackage{helvet}         
\usepackage{courier}        
\usepackage{type1cm}        
\usepackage{makeidx}         
\usepackage{graphicx}        
\usepackage{multicol}        
\usepackage[bottom]{footmisc}

\graphicspath{{Figures/}}
\DeclareGraphicsExtensions{.pdf,.png,.jpg,.eps} 

\usepackage{quoting}

\newcommand*{\myquotingsource}{}
\newenvironment{myquoting}[1]{%
  \renewcommand*{\myquotingsource}{#1}
  \begin{quoting}[font={itshape},leftmargin=0.25\linewidth,rightmargin=2em]%
   \DeclareStringOption{begintext}{``}  
}{%
  \DeclareStringOption{endtext}{''} 
  
  \raggedleft\myquotingsource
  
  \end{quoting}%
}

\makeindex             

\begin{document}

\title*{Physically Unclonable Functions and AI }
\subtitle{\large Two Decades of Marriage*}
\author{Fatemeh Ganji and Shahin Tajik}
\institute{Fatemeh Ganji and Shahin Tajik\at
Worcester Polytechnic Institute, \email{\{fganji, stajik\}@wpi.edu}
\\\footnotesize{$^*$ This is the authors' version of a survey to appear in the proceedings of the workshop ``AI+Sec: Artificial Intelligence and Security'', which was held 2--6 December, 2019, in Lorentz Center (for more details, see \url{https://www.lorentzcenter.nl/aisec-artificial-intelligence-and-security.html} [Accessed February 11, 2021]).} This chapter also partially covers what has been presented in the tutorial ``Security of PUFs: Lessons Learned after Two Decades of Research'' given at CHES 2019. }
%
%
\maketitle
\begin{myquoting}{\cite{goldreich2007foundations}}
The design of cryptographic systems must be based on firm foundations, whereas ad-hoc approaches and heuristics are a very dangerous way to go. 
\end{myquoting}

\abstract*{The current chapter aims at establishing a relationship between artificial intelligence (AI) and hardware security. 
Such a connection between AI and software security has been confirmed and well-reviewed in the relevant literature. 
The main focus here is to explore the methods borrowed from AI to assess the security of a hardware primitive, namely physically unclonable functions (PUFs), which has found applications in cryptographic protocols, e.g., authentication and key generation. 
Metrics and procedures devised for this are further discussed. 
Moreover, by reviewing PUFs designed by applying AI techniques, we give insight into future research directions in this area.}

\abstract{
The current chapter aims at establishing a relationship between artificial intelligence (AI) and hardware security. 
Such a connection between AI and software security has been confirmed and well-reviewed in the relevant literature. 
The main focus here is to explore the methods borrowed from AI to assess the security of a hardware primitive, namely physically unclonable functions (PUFs), which has found applications in cryptographic protocols, e.g., authentication and key generation. 
Metrics and procedures devised for this are further discussed. 
Moreover, by reviewing PUFs designed by applying AI techniques, we give insight into future research directions in this area. 
}

\section{Introduction}\label{sec:introduction}
In order to realize a cryptographic protocol or primitive, the assumptions made during design must hold. 
These assumptions relate in particular to secure key storage and secure execution of the protocol/primitive, which have been proven hard to attain in practice. 
The notion of root-of-trust has been introduced to deal with this by providing adequate reasoning in relation to physical security, i.e., the guarantee to resist certain physical attack cf.~\cite{maes2013}. 
In this regard, physically unclonable functions (PUFs) have been identified as a promising solution to secure key generation and storage issues~\cite{gassend2002silicon,gassend2004identification}. 

\begin{figure}[t]
    \centering
     \includegraphics[width=0.25\textwidth]{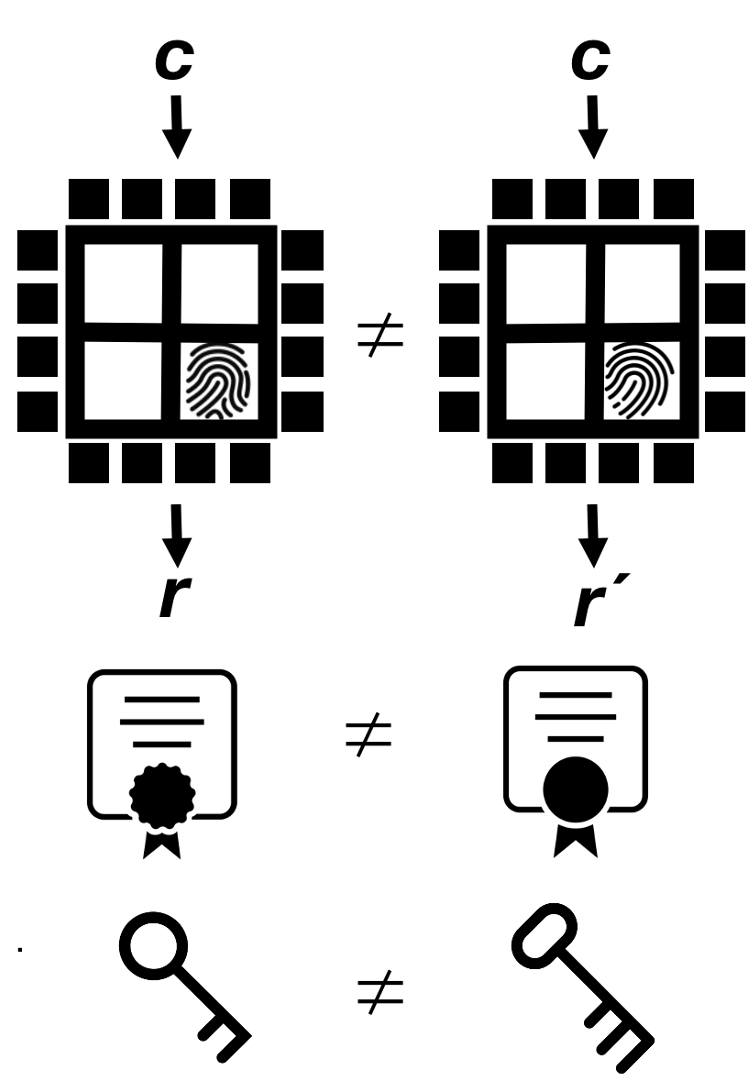}
     \caption{A conceptual illustration of a PUF: 
     For PUFs, the inevitable effect of manufacturing process variation is reinforced to enable PUF-based authentication systems and key generation. 
     To this end, the PUF is given an input (i.e., a challenge) to generate a (statistically) different output (so-called response) depending on the physical characteristic of the IC embodying the PUF. }
     \label{fig:PUFs_gen_concept}
\end{figure}

The key premise of PUFs is that physical characteristics of the system embodying the PUF can be tailored to derive an instance-specific feature of that, which is inherent and \emph{unclonable}~\cite{maes2013} (see Figure~\ref{fig:PUFs_gen_concept}). 
Among the variants of PUFs being considered in the literature are the so-called ``intrinsic'' PUFs, where the above feature (1) is the result of the production process, and (2) can be evaluated/ measured internally~\cite{guajardo2007fpga}. 
A prime example of PUF families meeting these criteria is (some types of) silicon PUFs with manufacturing process variations as the production process. 
One of the main advantages of these PUFs is the easy-to-integrate aspect, enabling the direct application of the PUF in integrated circuits (ICs).
This is of great importance for ICs used in not only every-day applications, but also cryptosystems. 

In the literature, a great number of research studies have been carried out to evaluate the properties of PUFs; at the heart of them is the unclonability. 
Obviously, if this property of a PUF is not fulfilled, neither the PUF nor the system encompassing that is secure.  
This chapter is devoted to the relationship between PUFs and AI, where the latter is used in a natural way to assess not only the security of PUFs, but also design PUFs. 
Among such approaches, if we focus on machine learning (ML)-enabled ones, the following interesting observation can be made.  

For PUFs, as cryptographic primitives, one of the most effective techniques to assess their security lies at the intersection of cryptography and machine learning, namely provable ML algorithms. 
In addition to exploring the difference between such algorithms and ML algorithms commonly employed in various fields of studies, this chapter describes why provable ML algorithms should be considered when evaluating the security of PUFs. 
For this purpose, pitfalls in methods adopted to demonstrate the robustness of a PUF against ML attacks are explained. 
Besides, the metrics that have been defined to assess this are mentioned. 

We put emphasis on the point that this chapter serves as neither an introduction to the concept of PUFs, nor their formalization and architectures. 
For these topics, we refer the reader to the seminal work and surveys published over the past decades, e.g.,~\cite{armknecht2011formalization,MaVe10,gao2020physical,maes2013}. 
\begin{figure}[t]
    \centering
     \includegraphics[width=1\textwidth]{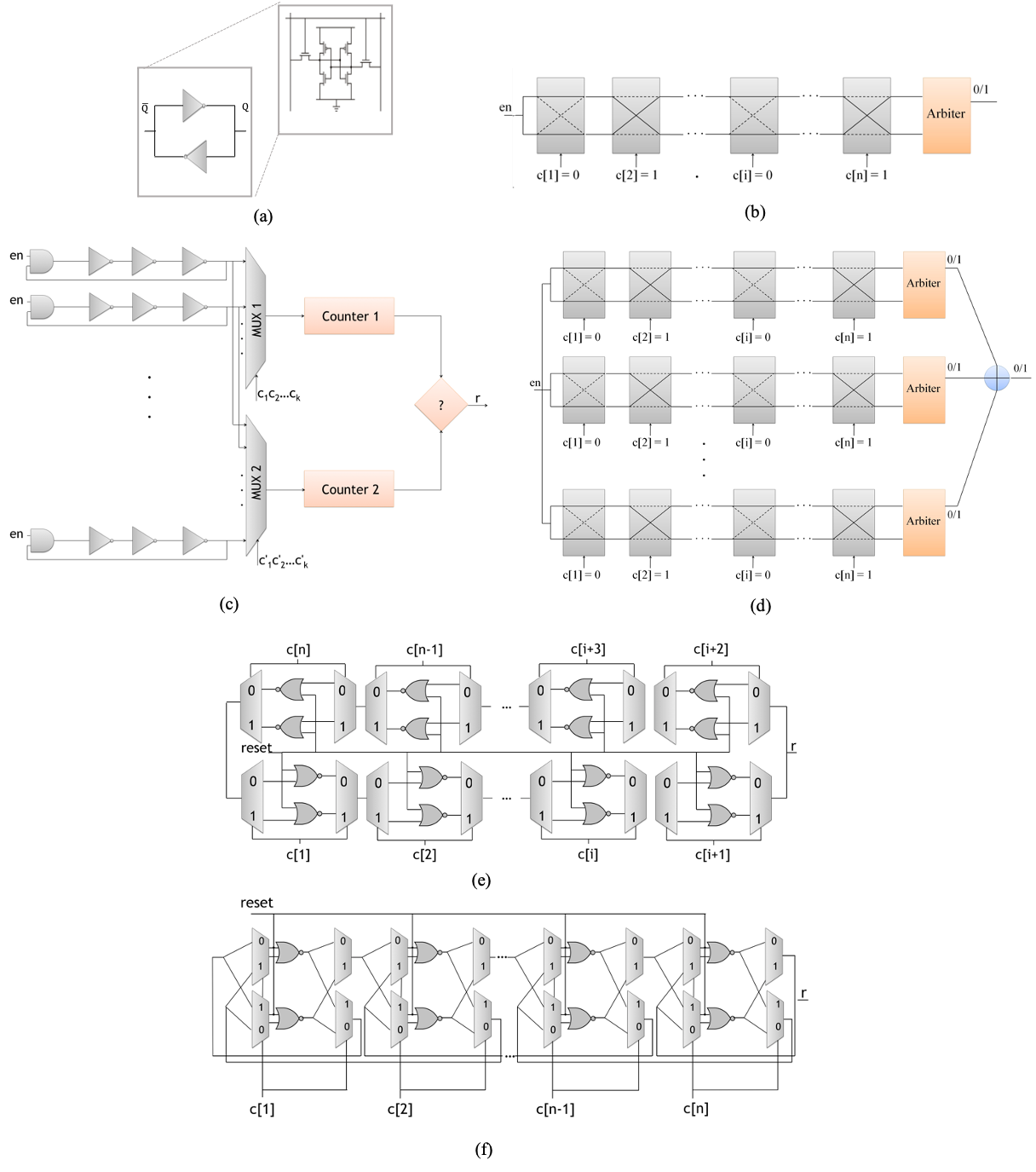}
     \caption{PUF Architectures. (a) A cell of SRAM PUF along with its CMOS circuit~\cite{maes2013}. (b) An Arbiter PUF. (c) A ring oscillator (RO) PUF. (d) An XOR Arbiter PUF with chains of Arbiter PUFs. (e) A bistable PUF (BR). (f) A twisted BR.}
     \label{fig:PUFs}
\end{figure}

\section{Background on PUFs}\label{sec:PUF_back}
PUFs exploit the process variations and imperfections of metals and transistors in similar chips to provide a device-specific fingerprint.
More formally, a PUF is a mathematical mapping generating virtually unique outputs (i.e., responses) to a given set of input bits (i.e., challenges).
These responses can be used either to authenticate and identify a specific chip or as cryptographic keys for encryption/decryption. 
To use a PUF in an authentication scenario, the PUF has to go through the \emph{enrollment} and \emph{verification} phases.
In the enrollment phase, the verifier creates a database of challenge-response-pairs (CRPs) for the PUF.

During the verification phase in the field, a set of challenges from the verifier is fed to the PUF, and the generated responses will be compared to the stored responses in the database for verification. 
In this phase, due to the natural noisy characteristics of PUFs, the verifier must resolve the noisy CRPs, i.e., for a given challenge, the response could be different, when the measurement is repeated (so-called the reliability problem). 
For this purpose, application of majority voting and \emph{fuzzy extractors} are the prominent examples~\cite{delvaux2016efficient}. 
The latter is a sub-set of helper data algorithms, performing post-processing to meet key generation requirements: reproducibility, high-entropy, and control. 
For this purpose, such algorithms generate helper data that can be stored in insecure (off-chip) non-volatile memory (NVM), for instance, or by another party~\cite{delvaux2014helper}. 
It is worth mentioning that the helper data is considered public. 
More importantly, this data inevitably leaks information about the PUF responses, leaving the door open for the cloning of PUFs. 

\paragraph{\textbf{PUF Instantiations}}
There are several ways to categorize PUFs.
One of the most well-known ways to classify them is based on the amount of CRPs that they can provide.
PUFs with a small number of CRPs are considered as weak PUFs, and PUFs with an exponential number of CRPs are considered strong PUFs.
The most prominent examples of weak PUFs are SRAM PUF, Ring-Oscillator (RO) PUF, and Butterfly PUF.
SRAM PUFs~\cite{holcomb2008power} exploits the random power-up pattern inside an  SRAM to generate a unique fingerprint.
Since the inverters inside each SRAM cell (Fig.\ref{fig:PUFs}(a)) have mismatches due to process variations, it is predictable at which logical value the cell is settled after power-up.
In this case, the challenge to the PUF can be the address of memory cells, and the response is the value stored in each cell.
The RO PUF, on the other hand, utilizes the intrinsic differences between frequencies of equal lengths ROs to produce the unique fingerprints.
In the case of RO PUF~\cite{suh2007physical}, a pair of ROs is selected by a given challenge, and the response is a binary value based on the comparison of the RO frequencies, see Fig.\ref{fig:PUFs}(c).

The underlying principles of the strong PUFs are very similar to weak PUFs, i.e., exploiting bistable and delay-based circuits.
However, in contrast to weak PUFs, the components of strong PUFs are tailored in a way that results in an exponential number of CRPs.
The most prominent strong PUF is the Arbiter PUF family~\cite{lee2004technique}, where the intrinsic timing differences of two symmetrically designed paths, chosen by a challenge, are exploited to generate a binary response at the output of the circuit, see Fig.\ref{fig:PUFs}(b).
It has become clear from the very beginning that the Arbiter PUF is vulnerable to machine learning attacks (see Section~\ref{sec:PUFs_attacks} for more details). 
Therefore, XOR Arbiter PUFs and other Arbiter-based PUFs have been proposed to mitigate the vulnerability of PUFs to machine learning attacks, see Fig.\ref{fig:PUFs}(d).
Further research in the area of strong PUFs has led to other strong PUFs constructions, which are inspired by bistable memory cells.
The main instances of these classes are bistable-ring (BR)~\cite{chen2011bistable} and twisted bistable-stable (TBR) PUFs~\cite{schuster2014evaluation}, see Fig.\ref{fig:PUFs}(e) and Fig.\ref{fig:PUFs}(f).
\begin{figure}[t]
    \centering
     \includegraphics[width=\textwidth]{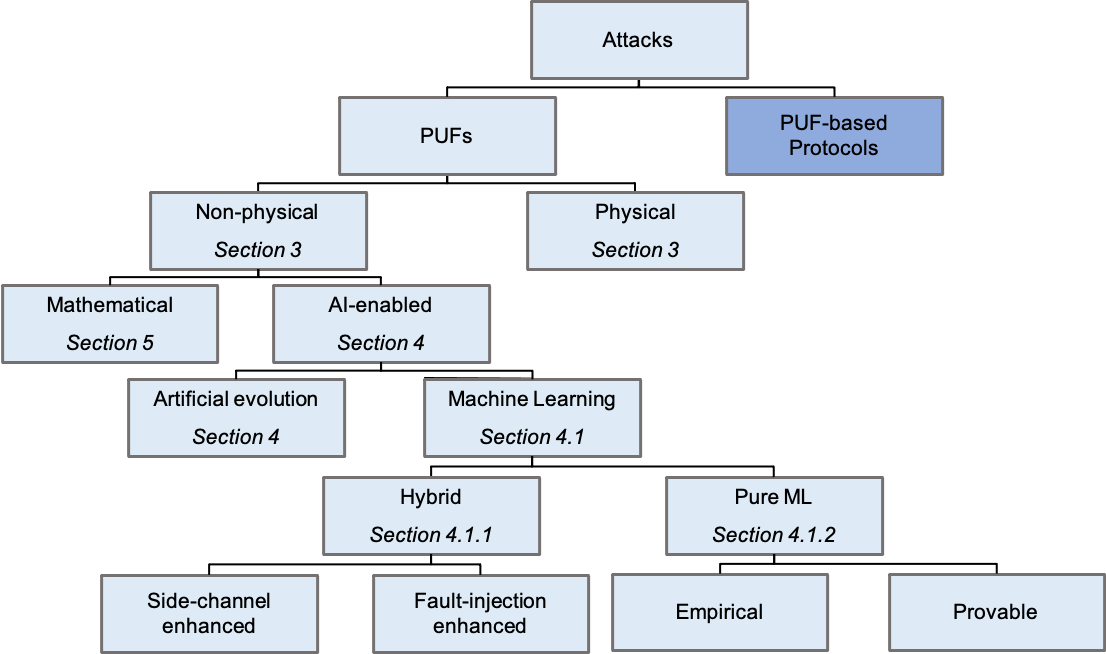}
     \caption{The taxonomy of attacks launched against PUFs.
     This chapter mainly focuses on AI-enabled, non-physical attacks; however, some studies on the physical and mathematical analysis of PUFs are briefly discussed. 
     Although attacks against PUF-based protocols are beyond the scope of this chapter, an interesting, AI-enabled example of such attacks is mentioned in Section~\ref{sec:puf_attacks_AI}. }
     \label{fig:PUF_attack}
\end{figure}
\section{Attacks against PUFs: Physical vs. Non-physical}\label{sec:PUFs_attacks}
In the same vein as other security primitives, designs have been coming hand in hand with attacks. 
Irrespective of the methods and means, through which an attack is carried out, the goal of the adversary is to determine how the inputs are mapped to outputs of the PUF, thereby predicting the responses to unseen challenges.
As broad as the range of PUF designs, attacks against PUFs have covered the whole spectrum in terms of different levels of destruction that they could cause. 
At one end of the spectrum, there are physical invasive attacks attempting to clone a PUF physically following a destructive procedure, e.g., through Focused Ion Beam (FIB) circuit edit\footnote{Note that through this chapter, to classify the attacks, the step taken to \emph{clone} a PUF is considered.}~\cite{helfmeier2013cloning}. 
While fully-invasive attacks can arbitrarily alter the functionality of the circuit and disable any countermeasures, semi-invasive attacks with a lower level of destruction can be launched to extract the ``secret'' from the PUF cf.~\cite{nedospasov2013invasive}. 
Here, secret extraction refers to the process of either reading-out the PUF responses~\cite{nedospasov2013invasive} or measuring some parameters that can be further analyzed to characterize a PUF~\cite{tajik2014physical,tajik2017photonic}. 
The latter is, of course, of greater interest to attackers as such attacks are not limited to memory-based PUFs. 
Interestingly, a group of semi-invasive attacks may not even damage the underlying PUF structure, see, e.g.,~\cite{merli2011semi}. 
Due to their nature, we categorize the above attacks as ``physical attacks,'' see Figure~\ref{fig:PUF_attack}. 

At the other end of the spectrum are physical non-invasive attacks, the adversary can observe/measure some challenge-response behavior-related parameters. 
The latter case is similar to that of side-channel attacks, widely studied in cryptography, and in particular, hardware security.  
Not surprisingly, PUFs have also been susceptible to side-channel attacks, which is not limited to memory- or delayed-based PUFs. 
An attempt has been made by the authors of~\cite{oren2013effectiveness} to present an attack taking advantage of remanence decay side-channels in SRAM PUF, which enables the attacker to recover the response of the PUF. 
Additionally, under the principle of side-channel analysis, an electromagnetic emission analysis against RO PUFs has been reported in~\cite{merli2013localized}. 

As the leakage from the helper data (see Section~\ref{sec:PUF_back}) can be thought of as a side-channel revealing some information regarding the PUF, attacks depending on this vulnerability are other particularly interesting examples of side-channel analysis. 
This type of attacks has been mounted against RO PUF constructions~\cite{delvaux2014key}. 
Similarly, the power leakage from a fuzzy extractor used to perform error correction has been exploited to recover PUF responses~\cite{karakoyunlu2010differential}. 
Another examples include attacks, where the measurements are fed into machine learning (ML) algorithms to predict the response of the PUF to unknown CRPs. 
Therefore, in our taxonomy, we classify these attacks into the ``AI-enabled'' category, and more concretely, in ``Machine Learning'' one, see Figure~\ref{fig:PUF_attack} (see Section~\ref{sec:PUF_attack_hybrid}). 

Next section covers a special type of non-invasive attacks, where AI - in particular, machine learning (ML) algorithms- enables the attacker to learn the challenge-response behavior of PUFs (see Figure~\ref{fig:PUF_attack}). 
These attacks either rely on the availability of some physical measurements (see Section~\ref{sec:PUF_attack_hybrid}) or perform learning solely based on available CRPs (see Section~\ref{sec:PUF_attack_pure}). 

\section{AI-enabled attacks}\label{sec:puf_attacks_AI}
This class is composed of attacks enhanced by incorporating techniques borrowed from the field of AI. 
In this regard, the application of ML algorithm has become a prominent subject for research due to their availability and ease of use. 
Nonetheless, other AI methods could be beneficial, when it comes to the security assessment of PUFs. 
For PUFs,~\cite{ruhrmair2010modeling,saha2013model} are two of the first studies focusing on this by employing evolution strategies and genetic programming against feed-forward Arbiter PUFs and RO PUFs, respectively. 
In this context, the security of recently emerging PUF architectures, e.g., non-linear current mirror PUFs~\cite{kumar2014design} has been compromised through genetic algorithms~\cite{guo2016efficient}. 

Further application of such algorithms has been observed in~\cite{becker2015gap}, where Becker et al. have applied an optimization technique relying on artificial evolution, in particular, covariance matrix adaptation evolution strategy (CMA-ES). 
In doing so, it has been shown that the security of an XOR Arbiter PUFs embedded in a real-world radio-frequency identification (RFID) tag can be compromised. 
Interestingly enough, to launch the attack, the reliability information\footnote{Repetition of the measurement by giving the same challenge results in different responses.} collected from the PUF, which can be thought of as a ``side-channel'', has been further exploited.  
A similar concept has been adopted against Feed-forward Arbiter PUFs and leakage current-based PUFs. 
However, fault injection has been performed to assist the evolutionary algorithm applied against these PUFs~\cite{kumar2014hybrid,kumar2015side}. 
Regardless of the issues raised about the convergence of the evolutionary strategies, see, e.g.,~\cite{nguyen2017mxpuf}, its scalability, and reliance on the side-channels/injected faults, these attacks have attracted a great deal of attention in the hardware security community. 
For instance, in addition to above-mentioned attacks against PUFs as hardware primitives,~\cite{delvaux2019machine} has demonstrated that recent Arbiter PUF-based protocols are susceptible to attacks employing the CMA-ES algorithm.

Next, attacks have been explained that apply the concept of \emph{learning} the challenge-response behavior of a PUF through ML algorithms (i.e., a subset of AI). 

\subsection{Machine Learning attacks}\label{sec:puf_attacks_ML}
As explained before, the main goal of the attacker is to mimic the challenge-response behavior of a PUF, preferably at a minimal cost. 
For this purpose, given the CRPs exchanged between the verifier and the chip, it is natural to think of ML attacks attempting to learn the mechanism underlying the response generation. 
In the early stages of introducing PUFs it was assumed that such attacks could not be successful~\cite{gassend2002silicon}; however, soon after conjecturing that PUFs are hard to learn, it has been experimentally verified that learning Arbiter PUFs is indeed possible~\cite{gassend2004identification}. 
This result initiated a line of research, being pursued for almost two decades now and resulted in studies covering various types of ML algorithms applied against a wide variety of PUFs. 
Among the ML algorithms discussed in the PUF-related literature are empirical and provable techniques, which we briefly review in this section. 
Before elaborating on \emph{pure} ML attacks, which are not refined by means of side-channels, we focus on hybrid, physical-ML attacks. 

\subsubsection{Hybrid Attacks}\label{sec:PUF_attack_hybrid}
The basic premise, on which these attacks are based, is the presence of physical, measurable quantities that either can enhance the success rate of a ML attack against a PUF or, from another perspective, need to be analyzed through ML. 
While the latter is being widely studied in other hardware-security domains, e.g., side-channel analysis against cryptosystems, the former has become more popular for PUFs.
This has been partially motivated by the existence of PUFs, which have been considered harder to learn. 
In this context, for XOR Arbiter PUFs, it has been proposed to incorporate the timing and power information as side-channels in order to improve the learning process~\cite{ruhrmair2014efficient}. 
Another example of such attacks has been explained in~\cite{becker2014active}, where controlled PUFs composed of an Arbiter PUF and the lightweight block cipher PRESENT has come under power side-channel analysis. 
The practical feasibility of these attacks has been questioned as accurate timing and power measurements may not be available.  

This problem can be overcome by combing the ML and fault injection attacks, although at the cost of being (semi-) invasive in some cases. 
XOR Arbiter PUFs have been attacked following this observation, where the fault is injected in a semi-invasive manner~\cite{tajik2015laser}. Nevertheless, fault injection procedures used to facilitate the learning process can be non-invasive as well; however, the famous examples of such attacks have not employed ML algorithms, but rather evolutionary strategies~\cite{becker2014active} and mathematical modeling~\cite{delvaux2014fault} (see Section~\ref{sec:PUF_attacks_math}).

\subsubsection{Pure ML Attacks}\label{sec:PUF_attack_pure}
Following the way paved by~\cite{gassend2004identification} to show the vulnerability of Arbiter PUFs to the Perceptron algorithm, Lim has demonstrated that a support vector machine (SVM) algorithm can also be used to learn the challenge-response behavior of Arbiter PUFs~\cite{lim2004extracting}. 
This line of research has been pursued and has led to exploring the application of two groups of ML algorithms in the context of PUFs, namely empirical and provable ML approaches. 
The former is widely taken in various domains, including hardware security, whereas the latter has been identified as a ``sister field'' of the cryptography and linked to complexity theory~\cite{rivest1991cryptography}. 
A major success factor of the provable techniques, in particular, Valiant's probably approximately correct (PAC) framework, is the link established between the probability of successful learning (so-called, confidence level), the number of training examples and their distribution, the complexity of the model assumed for the unknown function under test, and the accuracy of the learned model~\cite{valiant1984theory}. 
Such relationships cannot be made for empirical ML algorithms, more concretely, (1) the probability of successful learning is not known and cannot be defined prior to the experiments, and (2) the number of training examples cannot be determined beforehand. 
Moreover, empirical ML algorithms can suffer from a lack of generalizability and reproducibility. 
While generalization refers to the ability of the model to adapt properly to unseen data, drawn from the same distribution as of that followed to create the model, reproducibility means that an experiment can be repeated to reach the same conclusion. 

Although the above challenges are not specific to the security assessment of PUFs, this assessment may become impossible due to the lack of generalizability and reproducibility. 
To investigate the vulnerability of a PUF with instance-specific features and sensitivity to environmental changes, it is crucial to come up with ML approaches, making standardization and comparison feasible. 
For these purposes, provable ML algorithms seem promising. 
Next, a brief overview of empirical and provable attacks against PUFs is provided. 

\begin{table}[t]
\footnotesize
	 \caption{Some of the attacks enabled by empirical ML algorithm, with a focus on PUFs under attack, and models for representing the internal functionality of the respective PUFs. }
	  \centering
	  \begin{tabular}{p{0.09\textwidth}|p{0.32\textwidth}|p{0.32\textwidth}|p{0.25\textwidth}}
	    \hline
	    Ref. & PUF under attack   & ML algorithm  & Mathematical model \\
	    \hline
	    \cite{gassend2004identification} & Arbiter PUFs & Perceptron  & Linear combinations of delays\\
        \hline
        \cite{lim2004extracting} & Arbiter PUFs & Support Vector Machines (SVMs)  & Hyperplanes\\
        \hline
        \cite{RuSeSo10}& Arbiter PUFs, XOR Arbiter PUFs, Lightweight Secure PUFs & Logistic regression & Hyperplanes \\
        \hline
        \cite{RuSeSo10}&  RO PUFs & Quick Sort & NA \\
        \hline
        \cite{hospodar2012machine} & Arbiter and XOR Arbiter PUFs & Artificial Neural Networks and Support Vector Machines & NA\\
        \hline
        \cite{schuster2014evaluation} & Bistable Ring (BR) and twisted BR PUFs & Artificial Neural Networks & NA\\
        \hline
        \cite{xu2015security} & BR PUFs & SVM & LTFs\\
        \hline
        \cite{vijayakumar2016machine} & Non-linear voltage transfer characteristics (VTC) & Tree classifiers and  bagging and boosting techniques & NA\\
        \hline
        \cite{khalafalla2019pufs,awano2019pufnet} & Double Arbiter PUFs~\cite{machida2014new} & Deep learning & NA\\
        \hline
        \cite{santikellur2019deep} & XOR APUF, Lightweight Secure PUF, Multiplexer PUF and its variants & Deep learning & NA\\
        \hline
        \cite{wisiol2020splitting} & Interpose PUF~\cite{nguyen2019interpose} & Logistic regression & Hyperplanes\\
        \hline
	  \end{tabular}
	  \label{tab:PUF_attacks_emp}
	  \normalsize
\end{table}
\paragraph{Empirical ML Methods}
As mentioned before, along with the development of PUFs have come ML attacks~\cite{gassend2004identification}. 
These attacks have become more critical as Arbiter PUFs, XOR Arbiter PUFs, Lightweight Secure PUFs~\cite{majzoobi2008lightweight}, and RO PUFs have been successfully targeted by applying logistic regression~\cite{RuSeSo10}. 
This work has been followed by numerous studies that aim to assess the security of various PUF families against different empirical ML algorithms.
Table~\ref{tab:PUF_attacks_emp} summarizes some of these studies by focusing on PUFs and the ML algorithms discussed in those studies. 

\vspace{5pt}\noindent\textbf{Remarks: }
With regard to the attacks listed in Table~\ref{tab:PUF_attacks_emp}, an interesting observation can be made: except for some of the studies, no model, describing the functionality of the PUF, has been taken into account\footnote{Next, we discuss that for some provable techniques, this information is not required.}.  
Such models are viewed as ``data transformation'' in empirical ML-related literature. 
In the absence of these transformations, it is not straightforward to justify why an algorithm should be chosen to perform the learning task. 

Note that without such a justification, it is impossible to generalize the results obtained for a given PUF to other instances from that PUF family. 
This, of course, does not undervalue the importance of research on empirical ML algorithms in the context of PUFs; however, we emphasize that the results achieved through using such algorithms should be interpreted with caution. 

Another remark is that some of the PUFs listed in Table~\ref{tab:PUF_attacks_emp} were proposed as a remedy for ML attacks, but have been attacked by other algorithms. 
This raises the question of whether the resistance to ML should be defined as resistance to ``known ML attacks'', which is answered in Section~\ref{sec:PUFs_attack_resistance}. 
In line with this, another question is what to do to increase the security of PUFs targeted by ML attacks? 
An illustrative example is the XOR Arbiter PUFs, where adding non-linearity through using the XOR combination function is suggested.
Nevertheless, to understand the effectiveness of such a countermeasure, it is \emph{not enough} to rely on specific ML algorithms. 
Moreover, it is important to understand how and why ML algorithms could break the security of a PUF. 
Next, we discuss this in further detail. 

\begin{figure}[t]
    \centering
     \includegraphics[width=\textwidth]{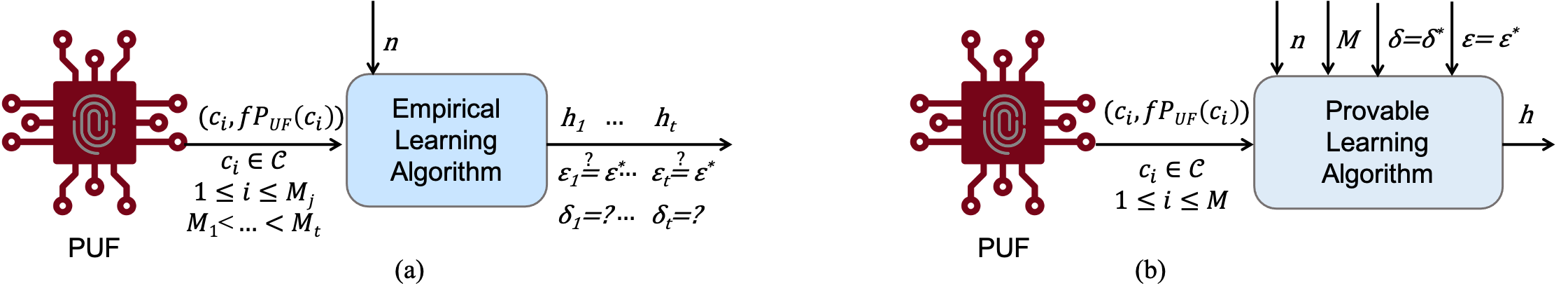}
     \caption{Schematic illustrating the differences between a provable and an empirical ML algorithm applied in the context of PUFs cf.~\cite{ganji2019pufmeter}. 
     For an $n$-bit PUF, $h$ denotes the model learned by the machine. $\varepsilon^*$ is the desired minimum accuracy of $h$, whereas the acceptable confidence level $\delta^*$ is the probability of obtaining the desired model. 
     On the contrary, for empirical ML algorithms, the latter parameter cannot be adjusted: it is not known if the algorithm converges to a model with the desired accuracy. 
     In fact, after each learning phase, e.g., the $j^{th}$ round $(1\leq j \leq t)$, the accuracy of the model $\varepsilon_j$ could be less than  $\varepsilon^*$. 
     Moreover, for provable ML algorithms, the maximum number of CRPs required for learning ($M$) can be computed as a function of $\varepsilon^*$, $\delta^*$, which is not the case for empirical ML algorithms.}
     \label{fig:PUF_provable_vs_empirical}
\end{figure}

\paragraph{Provable ML Methods}
What can be understood from the above discussion is that attacks mounted by applying empirical ML algorithms depend largely on trial-and-error approaches. 
The importance of this problem becomes apparent as countermeasures designed to defeat them would also render ineffective. 
Moreover, since security assessment with regard to empirical ML algorithms is instance-, parameter-, and algorithm-dependent with no convergence guarantees, standardization and comparison between PUFs may not be feasible. 
In response to this, a provable ML framework, namely probably approximately correct (PAC) learning, has found application in studies on PUFs. 
Figure~\ref{fig:PUF_provable_vs_empirical} presents the differences between an empirical algorithm and a provable one. 
It is worth noting here that ML attacks reported in the literature usually achieve an acceptable level of accuracy (e.g., 95\%) regardless of employing an empirical or a provable algorithm. 
Hence, the main advantage of provable algorithms is not the accuracy of the model learned, but the possibility of defining the level of accuracy and confidence a priori. 

PAC learning framework has been employed to launch provable ML attacks against PUFs. 
Families of PUFs targeted by PAC learning attacks include Arbiter~\cite{ganji2016pac}, XOR Arbiter~\cite{ganji2015attackers}, RO~\cite{ganji2015let}\footnote{Note that although weak PUFs are not interesting targets from the point of view of ML attacks, their characteristics can be analyzed by applying provable methods and the metrics defined based on them cf.~\cite{ganji2019pufmeter}.}, and BR and twisted BR PUFs~\cite{ganji2016strong,ganji2017having}. 
These results have been extended to the cases where the noisy CRPs are available to learn a PUF~\cite{ganji2018fourier}. 

PAC learning of BR and twisted BR PUFs serves as a special example since no mathematical model describing their internal functionality has been known so far; hence, no data transformation, or so-called ``representation'' could be presented. 
Regardless of that, characteristics of these PUFs are formulated as Boolean functions, useful to launch the attack. 
In this case, it has been proven that, in general, the challenge bits have different amounts of influence on the response of a PUF to a given challenge.  

Moreover,~\cite{ganji2017having} has reintroduced property testing in the PUF-related literature. 
Property testing algorithms developed in ML theory examine whether properties of a specific class can be found in a given function, i.e., a PUF under test in our case. 
In doing so, if no representation is known for a PUF, it is at least possible to understand whether it can be represented by some functions. 
It could be a necessary step when looking for an appropriate, efficient ML algorithm to learn a PUF. 
Furthermore, if an incorrect representation is chosen without either knowing the internal functionality of a PUF or performing property tests, it is not possible to decide about the learnability of the PUF~\cite{ganji2020pitfalls}.

\vspace{5pt}\noindent\textbf{Remarks: } We stress that when interpreting the above results, a great deal of attention should be paid to the setting of the PAC learning framework (access granted to the CRPs, distribution of the CRPs given to the machine, etc.) considered in a study. 
As explained in~\cite{ganji2020pitfalls}, this setting plays an important role when deciding if a PUF is provably (in)secure (see Section~\ref{sec:PUFs_attack_resistance} for more details). 

\section{Mathematical Modeling}\label{sec:PUF_attacks_math}
Mathematical modeling attacks do not fall within the category ``AI-enabled'' attacks, which is the main focus of this section; however, valuable lessons can be learned from these attacks. 
In this respect, one of the key messages to convey is that not every attack needs to be conducted with a view to adopting AI. 
In other words, security assessment of PUFs relying on AI, albeit a useful and important objective, should not be overstated as, in some cases, other mathematical methods can give more insight into the security of PUFs. 
This has been observed in a recent work of Zeitouni et al.~\cite{zeitouni2018s}, demonstrating that standard interpolation algorithms can be employed to predict the responses of (Rowhammer) DRAM-PUFs~\cite{schaller2017intrinsic}. 
Another instance is the analysis performed in~\cite{delvaux2013side} that has not only resulted in a high-accuracy model of the PUF under test, namely Arbiter PUFs, but also linked the reliability of a PUF to CMOS (and interconnect) noise sources and further to the PUF model. 
As shown for Arbiter, RO and RO sum PUFs, the effectiveness of this attack can be enhanced by injecting fault through changes in environmental conditions~\cite{delvaux2014fault}. 

Another example of such attacks concerns the so-called ``cryptanalysis'' of PUFs, where the adversary mounts computational attacks to predict the response to an unseen challenge with the probability higher than that of a random guess~\cite{sahoo2015case,nguyen2015efficient}. 
For this purpose, she can use the CRPs to partition the challenge-response space into subsets corresponding to the responses output by the PUF. 
To evaluate the effect of their attacks, the authors have targeted and successfully broken the security of RO PUFs with enhanced CRPs~\cite{maiti2011robust}, composite PUFs~\cite{sahoo2014composite} and the lightweight secure PUF~\cite{majzoobi2008lightweight}, with multi-bit responses. 

Within this category, the analysis presented in~\cite{yamamoto2014security} plays a crucial role in understanding the mechanism of generating responses in BR PUFs. 
More specifically, by carrying out linear and differential cryptanalysis-inspired attacks, this study has shown that the response of a BR PUF can be determined by a few number of the challenge bits. 
This work motivates the study of provable techniques to quantify the influence of challenge bits on the responses of PUFs~\cite{ganji2016strong,ganji2019pufmeter}. 

Finally, the attack presented in~\cite{ganji2015lattice} can be considered, where a lattice basis-reduction method has been applied to XOR Arbiter PUFs with an arbitrarily large number of chains and controlled inputs and outputs, i.e., unknown challenges and responses from the adversary perspective. 
These methods have been first found in a number of cryptography problems, e.g., hidden subset sum problem (i.e.,a variant of the subset sum problem with the hidden set of summands). 
The similarity between this problem and controlled XOR Arbiter PUFs can become evident as the side-channel analysis is conducted to measure the accumulated delays at the output of the last stage of the Arbiter PUFs in the XOR Arbiter PUF. 
It has been demonstrated that this information is sufficient to reveal not only the hidden challenges and responses, but also model the PUF. 

\vspace{5pt}\noindent\textbf{Remarks: }Perhaps, the most important message of this section is that the failure of AI-enabled attacks per se should not be considered as a guarantee for the security of PUFs. 
As explained above, PUFs that are robust to some AI-enabled attacks, e.g., XOR Arbiter PUFs with a large number of chains, can be vulnerable to other families of attacks. 
These attacks can, of course, improve designers' understanding of mechanisms underlying the challenge-response behavior of PUFs, and consequently, results in developing more attack-resilient PUFs. 

\section{Resiliency against ML attacks}\label{sec:PUFs_attack_resistance}
The attacks discussed in the previous section have posed serious challenges for the PUF designers and manufacturers. 
To tackle this issue, various countermeasures have been introduced in the literature, including controlled PUFs~\cite{majzoobi2009techniques,yu2016lockdown}, re-configurable PUFs~\cite{majzoobi2009techniques}, PUFs with noise-induced CRPs~\cite{yu2014noise,yashiro2019deep,wang2019adversarial}, to name a few. 
``Controlled PUFs'' is the umbrella term given to PUFs, where the adversary has restricted access to the CRPs through either obfuscating the challenges/responses~\cite{gassend2002controlled,gassend2008controlled} or mechanisms used to feed the challenges/collect the responses~\cite{majzoobi2009techniques,yu2016lockdown,lao2011novel,ye2016rpuf,katzenbeisser2011recyclable,lao2011novel,dubrova2019crc}. 
When it comes to re-configurability, PUFs with a mechanism to update the architecture of a PUF are devised~\cite{kursawe2009reconfigurable,gehrer2014reconfigurable,gehrer2015using,sahoo2017multiplexer}. 

When reviewing the papers mentioned above and in this section in general, it becomes evident that virtually all of the PUFs proposed in the literature have been designed having ML attacks in mind. 
This seems, however, paradoxical as attacks against PUFs, supposed to be ML attack-resilient, are being reported, see, e.g.,~\cite{delvaux2019machine,wisiol2020splitting}. 
The reason behind this can be the lack of (1) procedures to \emph{prove} that the PUF exhibits this feature, and (2) metrics to assess whether a PUF is robust against ML attacks.   

\subsection{How to Prove the Security of a PUF against ML Attacks}\label{sec:PUF_resiliency_proof_security}
Proving the security of cryptographic primitives has been practiced for decades. 
For that, the security of the system must be defined, which is typically carried out through the definition of an adversary model and a security game.
The former determines the power of the adversary in terms of, e.g., being uniform/non-uniform, interactive/non-interactive, etc. 
Adversary models also describe how the attacker interacts with the security game. 
This game further gives insight into the power of the adversary over the cryptosystem, i.e., her access to the systems and the conditions for considering an attack successful. 

Definition of security from the perspective of information theory (e.g., ``perfect secrecy'') and computational complexity have been employed to argue about the security of cryptosystems. 
This is natural as an adversary (a.k.a., ``codebreaker'') has bounded computational resources. 
Consequently, designers attempt to make the codebreaking problem computationally difficult to ensure security. 
This definition of security in cryptography has been linked to machine theory in a seminal work of Rivest in 1991~\cite{rivest1991cryptography}. 
Rivest observed the reliance of cryptography and machine learning on computational complexity, and further identified the similarities and differences between these two fields of study, including attack types and the queries required by an ML algorithm, exact versus approximate inference of an unknown target, etc.  

To assess the security of a scheme against a learning algorithm, i.e., the adversary, the ML setting is specified in the PAC learning framework. 
In this regard, one can define a set of parameters including the number of examples given to the algorithm, the distribution from that they are drawn (if needed), a representative model of the target function (if any), and the accuracy of the approximation. 
Afterward, if \emph{any} polynomial algorithm can learn the target function describing the cryptosystem, the security of the system is compromised. 
For PUFs, as cryptographic primitives, the same procedure can be followed; however, approaches proposed in the PUF-related literature have not pursued this. 

In particular, it is common that the robustness of a newly designed PUF against ML attacks is evaluated by applying a couple of ML algorithms against the PUF.  
This is indeed not sufficient to claim that the PUF is robust against ML attacks in general. 
In the best case, if the experiment is repeated for numerous instances of a PUF to make the results statistically relevant, one can claim that the PUF is resilient to the specific ML algorithm applied in the experiment. 

In contrast to such an ad-hoc process, the security of PUFs in the face of ML attacks have been analyzed rigorously in~\cite{ganji2020rock,hammouri2008tamper,yu2016lockdown,herder2016trapdoor}. 
Specifically,~\cite{ganji2020rock} has dealt with this from the point of view of PAC learning and computational complexity. 
The security of a PUF proposed in~\cite{yu2016lockdown} has been based upon a result previously obtained with regard to the PAC learning framework~\cite{ganji2015attackers}. 
Herder et al.~\cite{herder2016trapdoor} have also designed their PUF on the basis of a reduction to a problem known to be hard to learn, namely learning parities with noise. 

We sum up by quoting from Shannon's work that has argued against imprecise proof of security.  

\begin{svgraybox}
``In designing a good cipher [...] it is not enough merely to be sure none of the standard methods of cryptanalysis work– we must be sure that no method whatever will break the system easily. This, in fact, has been the weakness of many systems. [...] The problem of good cipher design is essentially one of finding difficult problems, subject to certain other conditions''~\cite{shannon1949communication} (see also~\cite{arora2009computational}). 
\end{svgraybox}

\subsection{Metrics for Evaluating the Security of a PUF against ML attacks}\label{sec:PUF_resiliency_metric}
In order to verify the robustness of a PUF against ML attacks in practice, a set of metrics should be provided that is well suited to various ML attacks. 
This task has been already accomplished for the purpose of verification that a PUF exhibits features relating to the quality.
In this regard, a comprehensive set of metrics has been developed, which comprises cost, reliability, and security~\cite{kim2010statistics,maiti2013systematic,kang2012performance}. 
Nevertheless, metrics associated with the ML attacks have not been well studied in the literature. 
In the following paragraphs, some of such metrics proposed in the literature are discussed. 

\vspace{5pt}
\noindent\textbf{Entropy: } One of the first attempts to define a metric indicating the ML attack-resiliency has been made in~\cite{majzoobi2008testing,majzoobi2009techniques}. 
The authors have used entropy to quantify the unclonability, i.e., being resistant to reverse engineering and modeling (i.e., mainly non-physical attacks). 
Entropy of the PUF responses can be estimated by the uniqueness, or measured by inter-distance~\cite{kim2010statistics}. 
In more concrete terms, unclonability due to the ML attacks is translated to statistical prediction and measured by the Hamming distance between two challenges~\cite{majzoobi2008testing}. 

\vspace{5pt}
\noindent\textbf{Strict avalanche criterion (SAC): }
It has been suggested that unpredictability can be achieved if the SAC property is satisfied: by flipping a bit in the challenge, each of the output bits flips with a probability of one half~\cite{majzoobi2009techniques}. 
This has been further improved in~\cite{nguyen2016security}, where the authors have shown that the distance between two challenges considered for computing the metric plays an important role. 

\vspace{5pt}
\noindent\textbf{Bias: } In addition to the above metrics, it is obvious that the bias in the PUF responses can be beneficial to launch mathematical, in particular, machine learning attacks.

\vspace{5pt}
\noindent\textbf{Noise sensitivity: }
The study in~\cite{nguyen2016security} has presented a prime example illustrating the lack of firm foundations for evaluating of resiliency of PUFs to ML attacks. To address this issue, as provable ML methods applied against PUFs have become established, new metrics originating from Boolean and Fourier analyses have been introduced~\cite{ganji2019pufmeter}. 
The notion of noise sensitivity is closely-related to the SAC property: the probability of flipping the output bit, when filliping each bit of the challenge with a pre-defined probability. 
It has been proved that the smaller and bounded the noise sensitivity is, the higher the probability of learning the PUF would be. 
More importantly, the algorithms for testing PUFs in terms of this metric have been made available cf.~\cite{ganji2019pufmeter}. 

\vspace{5pt}
\noindent\textbf{Expected bias: }Recent developments in the design and verification of PUFs with reliance on computational complexity result in adding a new metric to the list of formally-defined metrics: the expected bias~\cite{ganji2020rock}. 
The notion of the expected bias encompasses the average bias of a PUF in the presence of the noise, inherent to the design of the PUF due to, e.g., the effect of the routing, and/or having not sufficient deviation in the manufacturing process variations. 
These phenomena not only affect the responses of a PUF, but also induce correlation between two instances of a PUF implemented on a platform. 
Hence, the expected bias is a suitable measure to assess the security of PUFs, even ones composed of some PUF instances, e.g., XOR PUFs.

\section{AI-enabled Design of PUFs}\label{sec:PUFs_ML_PUFs}
Up until this point, we have focused on the interplay between the AI and PUFs from one point of view: applying ML to assess the security of PUFs.  
From another perspective, it is interesting to explore how AI can be employed to design PUFs. 
In the same way that ML-enabled attacks have been classified, this type of PUF design can be categorized into empirical and provable methods. 
In the first category, a PUF proposed in~\cite{de2019design} can be considered, where weightless neural networks (WNNs) have been adopted to transform the challenges in the sense of controlled PUF. 
Another interesting example of such PUFs has been proposed in~\cite{chowdhury2020weak}, where a weak PUF relying on the concept of asynchronous reset (ARES) behavior is designed. 
For this, a genetic algorithm has been employed with a fitness function defined based on the physical parameters of the transistors involved in the PUF circuit. 
In their scenario, the genetic algorithm automatically outputs an optimal PUF design for a given load.

On the other hand, in the second category, methods have been established that rely on the impossibility of learning specific functions. 
These impossibility results have been usually formulated in the PAC learning framework. 
Note that the impossibility taken into account here is not due to the setting chosen for the problem, e.g., representation, but the problem is inherently hard to learn. 
For instance, Herder et al.~\cite{herder2016trapdoor} have proposed a PUF relying on the hardness of learning parity with noise (LPN) problem. 
This problem remains open in computational learning theory, although it seems that the LPN problem is not as hard as NP-Hard problems~\cite{kalai2008agnostic}. 

Hammouri et al.~\cite{hammouri2008tamper} have presented one of the first studies on the application of PAC learning in the design of PUFs. 
For this, through reductions to problems that are known to be hard to learn, a proof of security has been suggested. 
This work has been followed by~\cite{yu2016lockdown}, where the design of the PUF is based upon a hardness problem proved for XOR Arbiter PUFs~\cite{ganji2015attackers}. 
The study presented in~\cite{ganji2020rock} has demonstrated that a family of functions, known to be hard-to-predict (i.e., Tribes function) can be applied to amplify the hardness of somewhat hard PUFs. 
Their proof of security against ML attacks relies on the unpredictability of the proposed PUF, composed of PUFs with lower levels of unpredictability. 

Interestingly, by adopting the concepts introduced in the context of PAC-learnability of PUFs, it has been shown that one could go one step further by devising an automated CAD framework to design a PUF, which can be provably ML attack-resilience~\cite{chatterjee2020puf}.  

\vspace{5pt}\noindent\textbf{Future directions: }Finally, we stress that compared to techniques originating in ML and used to attack PUF, not much effort has been invested in this topic. 
We expect that lessons learned from the AI-enabled attacks, and in particular, ML-based ones, become a tool for designing PUFs. 

\section{Conclusion}
This chapter attempts to demonstrate how AI can offer the potential for verifying that security-related features of a PUF are met. 
In this regard, the application of AI is not limited to processing the data to, e.g., classify the seen input-output pairs of a PUF and predict the output associated with an unknown input (i.e., mounting an ML attack). 
On the contrary, when shifting our focus to a provable ML framework, it is possible to come up with security proofs in the sense of cryptography. 
This has been explained in detail in this chapter, along with countermeasures, developed to protect PUFs against ML attacks. 
Moreover, metrics and approaches to quantify the robustness of a PUF to such attacks have been described. 
Last but not least, a new line of research devoted to the design of PUFs through ML techniques is discussed.

\begin{acknowledgement}
We are deeply grateful for the guidance and support of our Ph.D. advisor, Jean-Pierre Seifert. 
Throughout our Ph.D. studies at the Technical University of Berlin, and after that, his insightful comments and advice have helped us to broaden our knowledge and expand our horizons. 
We are also very thankful to Domenic Forte for his guidance,   tremendous support during our stay at the University of Florida, and encouragement, in particular, to make the tools (PUFmeter) publicly available.  

\end{acknowledgement}
\bibliographystyle{spmpsci}
\small
\bibliography{references}

\end{document}